\documentclass[preprint,12pt]{elsarticle}

\usepackage{graphicx}
\usepackage{float}
\usepackage{url}

\usepackage{setspace}
\usepackage{amssymb}
\usepackage{amsmath}
\journal{Proceedings of the Combustion Institute}

\begin{document}
%\small
%\baselineskip 10pt
\begin{frontmatter}


\title{Mechanism of flame acceleration and detonation transition from the interaction of a supersonic turbulent flame with an obstruction}
\author[add1]{Willstrong Rakotoarison}
\author[add2]{Brian Maxwell}
%\ead{email2@domain.ca}
\author[add3]{Andrzej Pekalski}
%\ead{email3@domain.ca}
\author[add1]{Matei I. Radulescu}
\ead{matei@uottawa.ca}

\address[add1]{Department of Mechanical Engineering, University of Ottawa, Ottawa ON Canada K1N 6N5}
\address[add2]{Department of Mechanical Engineering, University of Victoria, Victoria BC Canada V8W 2Y2}
\address[add3]{Shell Global Solutions, UK}

\begin{abstract}
The present paper seeks to determine the mechanism of flame acceleration and transition to detonation when a turbulent flame preceded by a shock interacts with a single obstruction in its path, taken as a cylindrical obstacle or a wall in the present study.  The problem is addressed experimentally in a mixture of propane-oxygen at sub-atmospheric conditions.  The turbulent flame was generated by passing a detonation wave through a perforated plate, yielding flames with turbulent burning velocities 10 to 20 larger than the laminar values and incident shock Mach numbers ranging between 2 and 2.5.  Time resolved schlieren videos recorded at approximately 100 kHz and numerical reconstruction of the flow field permitted to determine the mechanism of flame acceleration and transition to detonation. It was found to be the enhancement of the turbulent burning rate of the flame through its interaction with the shock reflection on the obstacle.  The amplification of the burning rate was found to drive the flame burning velocity close to the speed of sound with respect to the fresh gases, resulting in the amplification of a shock in front of the flame.  The acceleration through this regime resulted in the strengthening of this shock.  Detonation was observed in regions of non-planarity of this internal shock, inherited by the irregular shape of the turbulent flame itself.  Auto-ignition at early times of this process was found to be negligibly slow compared with the flow evolution time scale in the problem investigated, suggesting that the relevant time scale is primarily associated with the increase in turbulent burning rate by the interaction with reflected shocks. 
\end{abstract}


\end{frontmatter}
% \linenumbers
%% main tex
\section{Introduction}
\addvspace{10pt}
When a flame propagates through a congested area, the flame surface area increases as preferentially directed flow convecting the flame elongates its surface area.  This leads to an increase in the rate of reactant consumption, which in turn accelerates more fluid, thereby increasing the global burning rate.  This feedback mechanism leads to the acceleration of the flame front in the preferred direction of motion and the generation of a shock wave ahead of the flame.  These early dynamics are now very well understood through extensive experiments, theoretical models and numerics \cite{bychkov2008physical,kellenberger2015, Ciccarelli&Dorofeev2008,Lee&Moen1980}; accordingly the dynamics are also relatively well predicted at engineering scales of interest via coarse grained models advecting the surface of the flame, given a particular obstruction geometry (e.g., \cite{johansen2013modeling}).

The subsequent acceleration of the flame-shock complex and its possible transition to a detonation are not as well understood, and cannot be easily predicted \cite{oran2007origins, Ciccarelli&Dorofeev2008}.  The prime difficulty lies in the fact that the flame at this stage is no longer laminar \cite{kellenberger2015}.  Other than the mechanism described above, the turbulent burning rate itself at the flame becomes strongly coupled with the flow field generated by the flame and gasdynamic interactions \cite{oran2007origins}.  Turbulent transport and gas dynamic heating may both participate.  In the presence of obstacles, reflected shocks traversing the flame may promote turbulent mixing at the flame, leading to higher burning rates \cite{thomas2001experimental} conducive to a more prompt DDT. Experiments and numerical simulations usually agree that the final detonation formation involves the creation of hot-spots,  where  the Zel'dovich-Lee gradient mechanism \cite{Lee&Moen1980} can be set up in a variable number of ways: shock reflections on obstacles, turbulent mixing, self-generated shocks wave turbulence \cite{oran2007origins, Maleyetal2015, saif2017chapman}.  These complications make it difficult to predict the eventual transition to detonation, which is found to depend on many factors: e.g., turbulence intensity, flame properties, congestion geometry, sensitivity to auto-ignition, memory of the flow field set-up during the early stage of acceleration. 

The present study attempts to isolate the mechanism that is responsible for detonation transition of a shock-turbulent flame complex in the presence of an obstacle, in the hope that a criterion for flame acceleration and detonation transition can be identified.  We study the interaction of a well defined shock-flame complex, as it interacts with a single cylindrical obstacle or a flat wall.  Care is exercised to obtain a controllable initial turbulent flame of sufficient intensity, such that the subsequent transition to a detonation occurs rapidly and is reflective of the last stages of the DDT process.   The shock-turbulent flame complex in our study was generated using the technique of passing a detonation wave through a perforated plate, which gives rise to a system of interacting shocks yielding a region of intense wave turbulence.  This can give rise to an intense turbulent deflagration \cite{Chao2006, Grondin&Lee2010, Maleyetal2015} approaching the Chapman-Jouguet deflagration speed \cite{saif2017chapman, maxwell2018modelling} typically observed in fast flame propagation and DDT in tubes with constrictions \citep{Ciccarelli&Dorofeev2008}.  This permits to generate flames with high levels of turbulence conducive to transition to detonation from the interaction with a single obstruction, as we will show in the present study.

%%%%%%%%%%%%%%%%%%%%%%%%%%%%%%%%%%%%%%%%%%%%%%%%%%%%%%%%%%%%%%%%%%%%%%%%%%
\section{Experimental details}
\begin{figure}
\begin{center}
\includegraphics[width=140mm]{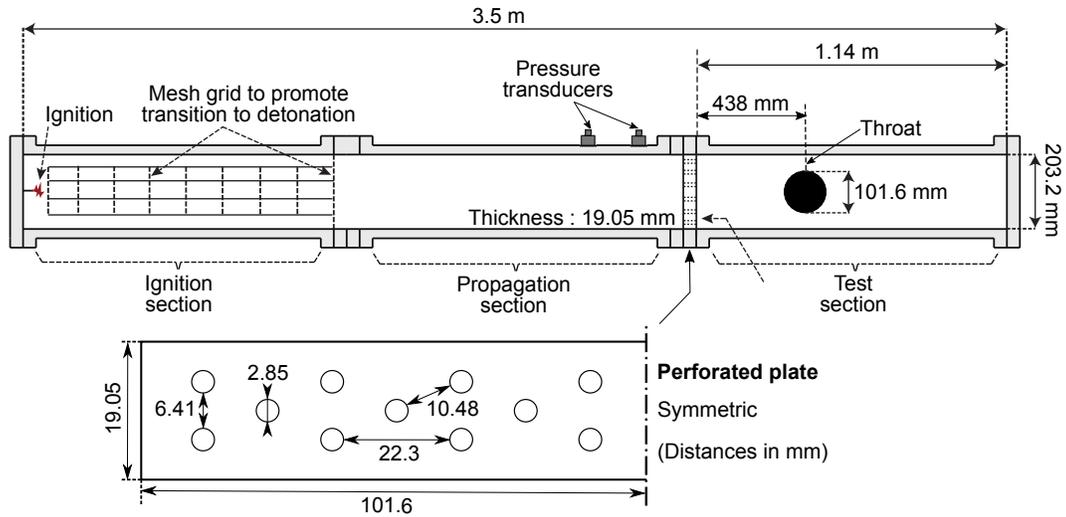}
\end{center}
\caption{Schematic of experimental set-up.}
\label{fig:schematic}
\end{figure}

\addvspace{10pt}
The experiments were conducted in a 3.5-m-long thin rectangular channel with a cross-section of 203 mm by 19 mm (Fig.\ \ref{fig:schematic}), described in detail elsewhere \cite{Bhattacharjee2013}. The entire tube was filled with the desired test mixture. A detonation was initiated by a high voltage capacitor discharge.  An initial obstacle section promoted this transition for less sensitive mixtures.  A self-sustained detonation wave was then established in the tube, which propagated at speeds approximately 5 \% below the Chapman-Jouguet value calculated.  This was inferred from time or arrival measurements using a pair of PCB 113B24 pressure transducers. The detonation then transmitted across a perforated plate as a leading shock followed by a turbulent deflagration.  The intensity of the turbulent deflagration and the strength of the leading shock were controlled by the perforated plate's blockage ratio, hole geometry and mixture sensitivity.  The configuration of the perforated plate investigated is shown in Fig.\ \ref{fig:schematic}.  The resulting high area blockage ratio of approximately 96\% was chosen higher than in our previous studies \cite{Maleyetal2015, saif2017chapman, maxwell2018modelling}, such that the strength of the transmitted shock was weaker, in order to de-couple the leading shock from the trailing flame, and the turbulent intensity higher.  

Two series of experiments are presented in this communication.  The first involves the interaction of the shock flame complex with a 101.6-mm-diameter cylinder, which was located 438 mm downstream of the perforated plate.  In the second series of experiments, the obstacle was removed in order to monitor the head-on interaction of the turbulent deflagration wave with the shock wave reflected by the end wall.  The reactive mixture studied was C$_3$H$_8$+5O$_2$, chosen such that it is sensitive enough at the low pressures required for the experimental facility available in our laboratory. The mixture sensitivity was changed by the initial pressure of the gas in the range of 4 to 10 kPa.  The ambient temperature was 294 K. The experiments were monitored with high speed schlieren visualization \cite{Bhattacharjee2013}.  The images acquired with a Phantom V1210 camera had a resolution of 384 by 288 pixels, corresponding to a spatial resolution of 0.8 mm per pixel.  All the experiments shown have an inter-frame time of 12.9 $\mu$s, typically providing approximately 100 useful frames per experiment. These sequential frames were then analyzed in order to reconstruct the flow-field, using additional theoretical and numerical tools, as described below.  
	 
%%%%%%%%%%%%%%%%%%%%%%%%%%%%%%%%%%%%%%%%%%%%%%%%%%%%%%%%%%%%%%%%%%%%%%%%%%
\section{Numerical tools}
\addvspace{10pt}
Partial reconstruction of the flow field measured experimentally relied on the Cantera thermo-chemical tools and the Shock and Detonation Toolbox developed by Shepherd and his students \cite{Cantera}.  Given the shock speeds measured in the experiments, the post shock states can be evaluated using the exact thermo-chemical properties of the reactants.  Likewise, laminar burning velocities and characteristic ignition delays can be reliably calculated using these tools at the various thermodynamic states of interest.  The current calculations were performed using the San Diego thermo-chemical database relevant for propane combustion \cite{Sandiego}.  

In some instances, partial numerical reconstruction of the two-dimensional \textit{inert} shock reflection process ahead of the flame was also performed in order to estimate the flow speed in the experiments used for extracting the local burning velocity.  The numerical calculations solved the inert Euler equations for a perfect gas with a compressible hydrocode developed by Falle \cite{falle1993body}, which uses an exact Godunov Riemann solver and adaptive mesh refinement.  All physical boundaries use a reflective boundary condition, while the back boundary used a zero gradient boundary condition.  The initial conditions correspond to a planar shock with uniform state behind it. A contact surface was imposed as initial condition at some distance behind the shock, with a density change corresponding to that of the flame in the experiments.  In these calculations, the incident shock Mach number was taken to match that determined in the experiment.  The specific heat ratio was taken as $\gamma$ = 1.22 evaluated using the method described above at the post-shock state in the experiments.  
%%%%%%%%%%%%%%%%%%%%%%%%%%%%%%%%%%%%%%%%%%%%%%%%%%%%%%%%%%%%%%%%%%%%%%%%%%%%%%%%%%%%%%%%%%%%%%%%%
\section{Results}
\addvspace{10pt}
\subsection*{Flame acceleration past a circular congestion}
Whether the interaction of a shock-flame complex with the cylinder resulted in the detonation initiation in the experiment was found to depend on the initial pressure of the mixture, which controlled the strength of the turbulent deflagration produced.  The effect of initial pressure on the frequency of detonation formation from the interaction was as follows.  At initial pressure of 4.8, 5.5, 6.2, 6.55 and 6.9 kPa, the frequency of detonation formation was respectively 0/9, 2/9, 3/4, 7/7 and 7/8.  Table \ref{table:parameters} lists the relevant physical properties of interest for the conditions bracketing this critical regime.  At each initial pressure, $M_I$ is the Mach number of the incident shock prior to its reflection on the cylinder, $S_u$ is the calculated laminar burning velocity with the unburned gases at the post shock state, $S_f$ is the experimentally measured burning velocity of the flame,  $T_{R}$ is the calculated temperature of the gases behind the normal reflected shock, while  $\tau$ is the ignition delay of the gases behind the reflected shock.  

\begin{table}
\begin{center}
   \caption{Thermo-chemical parameters for experiments shown}
   \begin{tabular}{l c c c c c c}
      \hline
                    &  $p_0$ (kPa) & $M_I$ & $S_u$ (m/s) & $S_f$ (m/s) & $T_{R}$ (K) & $\tau$ ($\mu$s)  \\
    \hline
      cyl. & 4.8 & 2.3 & 5.6 & 100 $\pm$ 24 & 653 & 3.9 $\times 10^5$\\
      cyl. & 6.9 & 2.5 & 6.1 & 140 $\pm$ 13 & 736 & 9.6 $\times 10^4$\\
    \hline
      wall & 4.1 & 2.0 & 5.0 & 56 $\pm$ 32 & 572 & 9.3 $\times 10^6$\\
      wall & 4.8 & 2.1 & 5.2 & 46 $\pm$ 46 & 605 & 1.9 $\times 10^6$\\
      wall & 6.2 & 2.2 & 5.4 & 210 $\pm$ 35 & 627 & 6.7 $\times 10^5$\\
      \hline
    \label{table:parameters}
    \end{tabular}\\
\end{center}
\end{table}

\begin{figure}
\begin{center}
\includegraphics[width=140mm]{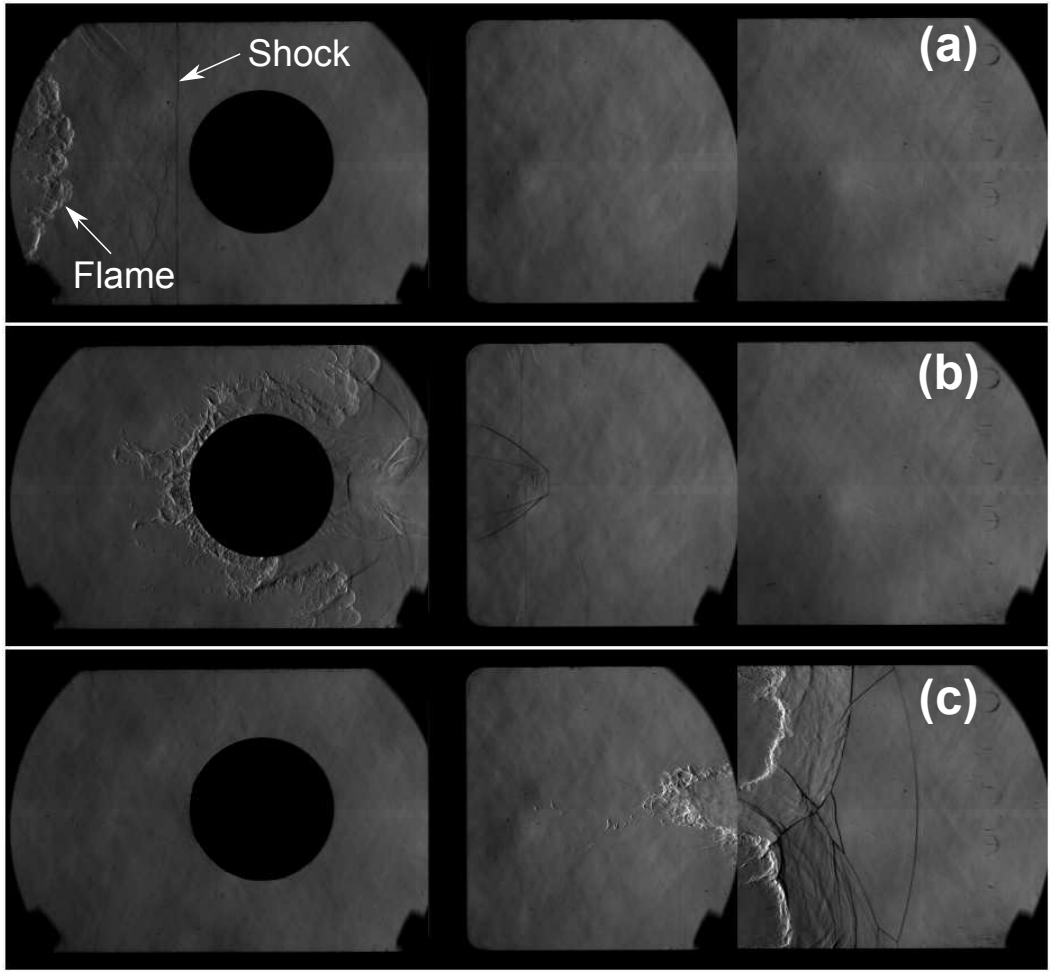}
\end{center}
\caption{The interaction of a turbulent flame-shock complex with a cylindrical obstacle in C$_3$H$_8$+5O$_2$ at 4.8 kPa initial pressure, composite images spliced from 3 different experiments; video animations as supplemental material; the time between successive frames is 425.7 $\mu$s.}
\label{fig:exp4-7-9circle}
\end{figure}

Figure \ref{fig:exp4-7-9circle} shows an example of a typical flow field evolution for a sub-critical initiation at 4.8 kPa initial pressure.  The evolution was obtained by splicing together three experiments, where the schlieren field was displaced progressively downstream of the cylinder.  Note the very good merging of the different experiments, which show the good reproducibility of the experiment.  In Fig.\ \ref{fig:exp4-7-9circle}a, the incoming shock flame complex is visible: a straight shock followed by a turbulent flame.  In Fig.\ \ref{fig:exp4-7-9circle}b, the leading shock has diffracted around the obstacle ahead of the flame.  The flame now has an extended surface, as the flow convects it around the cylinder.  In Fig.\ \ref{fig:exp4-7-9circle}c, the flame has acquired a V-shape, owing to its preferential convection around the cylinder.  Note the series of weak shock waves ahead of the flame, which were generated by the increase in global burning rate owing to the flame surface area increase.   In this case, we recover the well-known mechanism of flame acceleration and pressure wave generation due to the enlargement of flame surface area discussed in the introduction.  The video animation containing all consecutive frames is given as supplementary material.

\begin{figure*}
\begin{center}
\includegraphics[width=144mm]{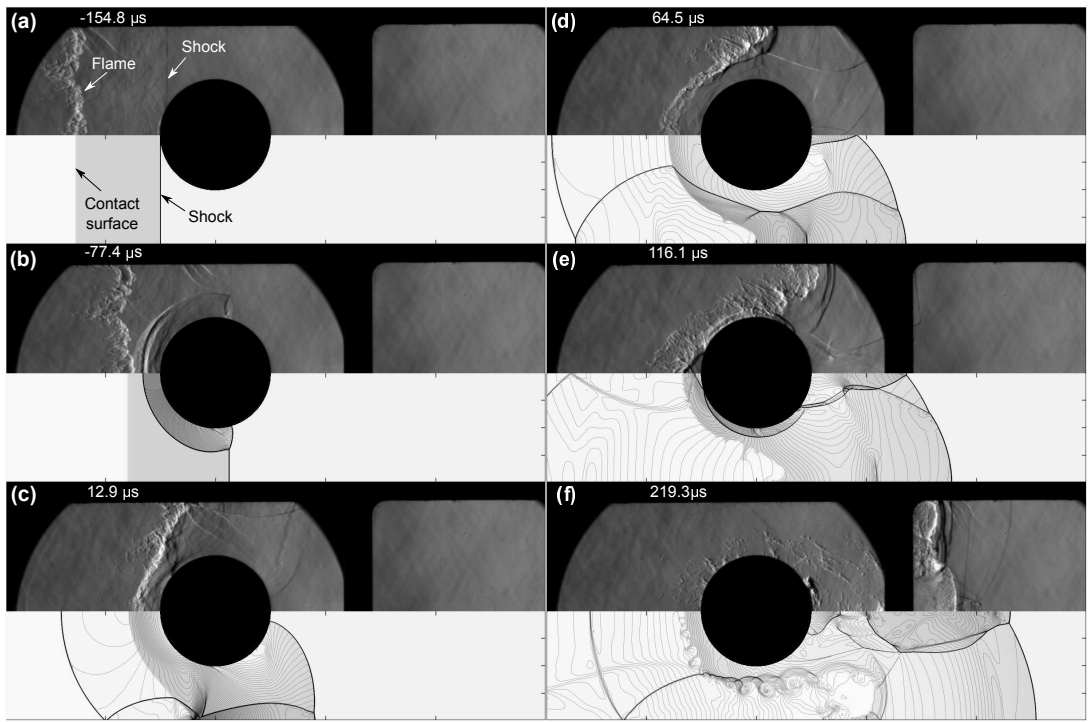}
\end{center}
\caption{The interaction of a turbulent flame-shock complex with a cylindrical obstacle in C$_3$H$_8$+5O$_2$ at 6.9 kPa initial pressure (top row in each frame) and the numerical reconstruction of an inert flow field (bottom row) showing a composite image of density field overlayed with pressure contours (video as supplemental material).}
\label{fig:exp26circle}
\end{figure*}

In sharp contrast, experiments above 6.9 kPa conclusively showed the prompt detonation initiation, with the detonation formation point moving closer to the point where the first reflected shock from the obstacle interacts with the turbulent flame. At the critical regime, the detonation initiation was sufficiently delayed to clearly reconstruct the dynamics.  Figure \ref{fig:exp26circle} shows the evolution of the initiation process at 6.9 kPa, when the detonation formation is delayed to the back of the cylinder. In the same figure, we show the numerical reconstruction of the \textit{inert} shock reflection process used to determine the local flow speed ahead of the flame. 

\begin{figure}
\begin{center}
\includegraphics[width=140mm]{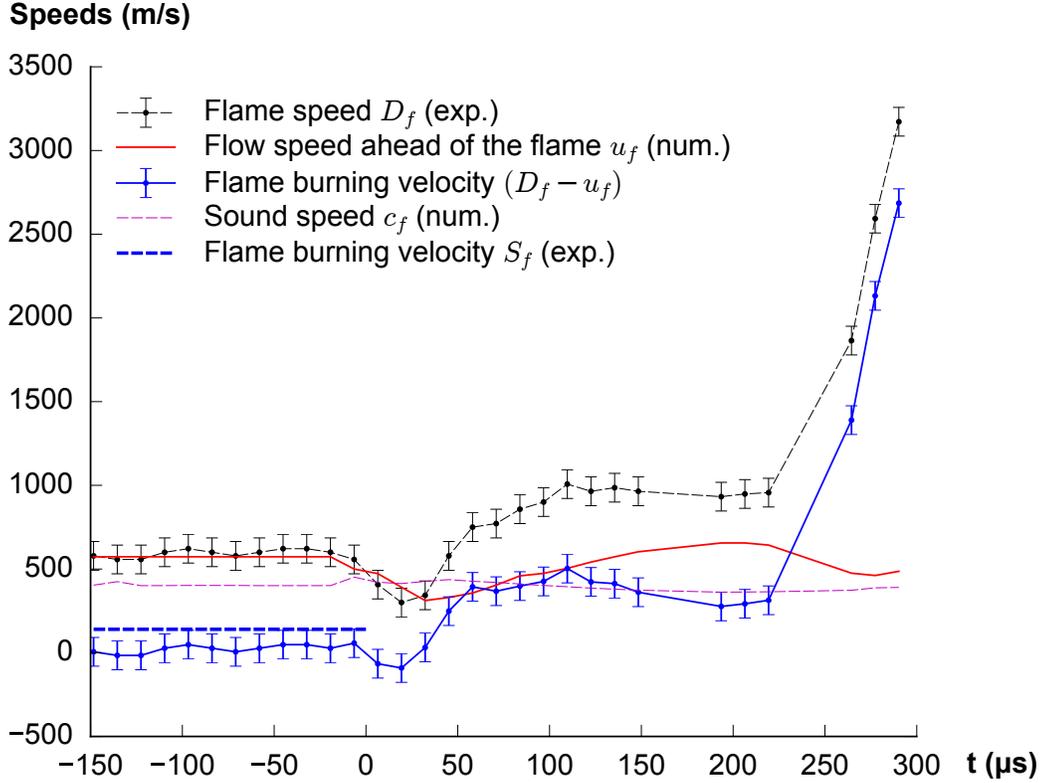}
\end{center}
\caption{The speed of the flame $D_{f}$, the flow speed ahead of the flame $u_{f}$ and the resulting flame speed relative to the inert flow ahead of it for the experiment of Fig. \ref{fig:exp26circle}; time 0 corresponds to the first interaction between the flame and the reflected shock.}
\label{fig:exp26circlespeeds}
\end{figure}

The combination of the experimental and numerical sequence first permits us to qualitatively comment on the flow field evolution. Figure \ref{fig:exp26circle}b shows the reflection of the leading shock on the cylinder.  In Fig. \ref{fig:exp26circle}c, the reflected shock has traversed the flame.  Up to this point, the inert calculation and the experiment are in qualitative agreement, since the local burning rate (initially approximately 140 m/s, see below) is relatively small compared to the convective fluid velocities (576 m/s behind the incident shock).  Substantial differences arise at later times in Figs.\ \ref{fig:exp26circle}d-f after the reflected shock has interacted with the flame.  In the experiments, the flame accelerates substantially, while the contact surface advected with the flow in the inert calculation  lags behind.  The flame acceleration in the experiments is seen to drive a shock wave ahead of it.  Note that a shock is there even in the inert calculation, its origin being the reflection of the reflected shock with the wall close to the flame surface (inert contact surface in the numerics) shown in Figs. \ref{fig:exp26circle}d. In the experiments, this internal shock-flame complex rapidly accelerates and transits to a detonation, either directly, or by subsequent shock reflections on the symmetry axis.  Both types of re-initiation were observed in the experiment illustrated in Fig.\ \ref{fig:exp26circle}.  
The sequential frames of Fig. \ref{fig:exp26circle} permit to reconstruct approximately the flame speed relative to the fresh gas moving ahead of the flame.  Analysis of the sequential frames in the experiment permits to determine the speed of the flame in laboratory coordinates $D_{f}$ along the top wall, shown in Fig. \ref{fig:exp26circlespeeds}. The reconstruction of the flow speed in front of the flame used two methods depending whether it was to estimate the flow speed ahead of the steady flame prior to the reflection or after the reflection.  Before reflection, the flame propagates into gas shocked by the incident shock.  Using the average shock speed, the gas speed immediately behind the leading shock was obtained by the jump conditions.  Since the flow behind the leading shock is not constant, owing to a weak deceleration of the leading shock in this case, a linear correction can be applied if the speed gradient was known.  This speed gradient behind the shock can be evaluated from first principles from the speed $D$ and decay rate of the leading shock $\dot{D}$ , yielding the so-called \textit{shock change} equation \cite{Fickett&Davis1979}: 
\begin{equation}
\frac{\partial u}{\partial x}=\frac{\left(1+\rho_0 D \left(\frac{d u}{d p}\right)_H\right)\frac{1}{\rho}\left(\frac{d p}{d D}\right)_H \dot{D}}{\left(D-u\right)^2 - c^2}
\end{equation}
where the pressure $p$, density $\rho$ and flow speed in the absolute frame $u$ are evaluated at the post shock state.  $\rho_0$ is the density of the undisturbed medium ahead of the shock and the subscript $H$ denotes a derivative taken along the shock Hugoniot.  In applying this relation, the deceleration of the leading shock $\dot{D}$ was measured experimentally by fitting a trajectory of constant acceleration to the experimentally determined $x(t)$ data.  The derivatives along the shock Hugoniot were evaluated numerically by perturbing the shock jump calculations by 1\% in shock speed.  The resulting burning velocity obtained is tabulated in Table \ref{table:parameters}.  The standard error reported for the burning velocity is associated with the error of the fitted parameter $\dot{D}$, obtained using the non-linear Levenberg-Marquardt least squares algorithm.  This error provides the largest source of error in our model estimate.  For this experiment, we obtain a turbulent flame burning velocity prior to the reflection of approximately 140 m/s, or approximately 22 times the laminar burning velocity calculated for this mixture at the post shock thermodynamic state.  

After the shock-flame interaction, the flame speed relative to the gases ahead was obtained by making use of the numerical calculations, which permitted to estimate the flow speed just ahead of the flame.  Indeed, since the flame develops initially a weak shock in the experiments, the region ahead of the shock is not affected by the flame, and the inert calculation provides a good measure for the local speed of the flow.  In the experiments, the flame speed was measured along the top wall, such that the flow speed (away from boundary layers) was always directed in the mean direction of motion.  The spatial and temporal evolution of these two speeds are shown in Fig.\ \ref{fig:exp26circlespeeds}.  While the absolute measured flame speed continuously increases, marking the transition to detonation at the end of the record, the advection of the flame by the flow is responsible for a large portion of this speed.  Indeed, the flow speed increase in the inert simulations is due solely to the acceleration of the flow passing through a narrower open region (the throat).  

The difference between the absolute flame speed and the calculated flow speed, shown in blue, is more meaningful.  It shows a rapid increase to approximately 400 m/s by the time the flame is at the throat.  Note that this speed corresponds closely to the local sound speed in the gas ahead of the flame. Past the throat, this effective turbulent burning velocity is slightly dropping, to only significantly re-accelerate once the flame has passed the cylinder. The view that emerges from this analysis is that the flame burning velocity is augmented by the shock flame interaction approximately three-fold. 

It is pertinent at this point to comment on the relative importance of auto-ignition at the early stages of this flame acceleration process.  Following the inert shock reflection, the temperature behind the reflected shock, before substantial volumetric expansion, is calculated from the exact shock jump conditions to be 736 K (see Table \ref{table:parameters}).  At this state, the constant volume ignition delay is approximately 100 ms.  This time scale is 3 orders of magnitude longer than the flow evolution and flame acceleration time scales apparent in Figs. \ref{fig:exp26circle} and  \ref{fig:exp26circlespeeds}, thus ruling out the mechanism of auto-ignition at a hot spot driven by shock reflection. The flame acceleration is thus due to the increased burning rate by the passage of the reflected shock. 

\subsection*{Flame shock head-on interactions}
In order to better determine the mechanism responsible for the rapid increase in burning velocity after the shock flame interaction, the second series of the experiments focused on the problem of normal reflected shock interaction with the flame. These series of experiments were performed by removing the cylinder from the shock tube, and allowing the incident shock to reflect on the end wall of the shock tube and interact with the flame following it.  

\begin{figure}
\begin{center}
\includegraphics[width=140mm]{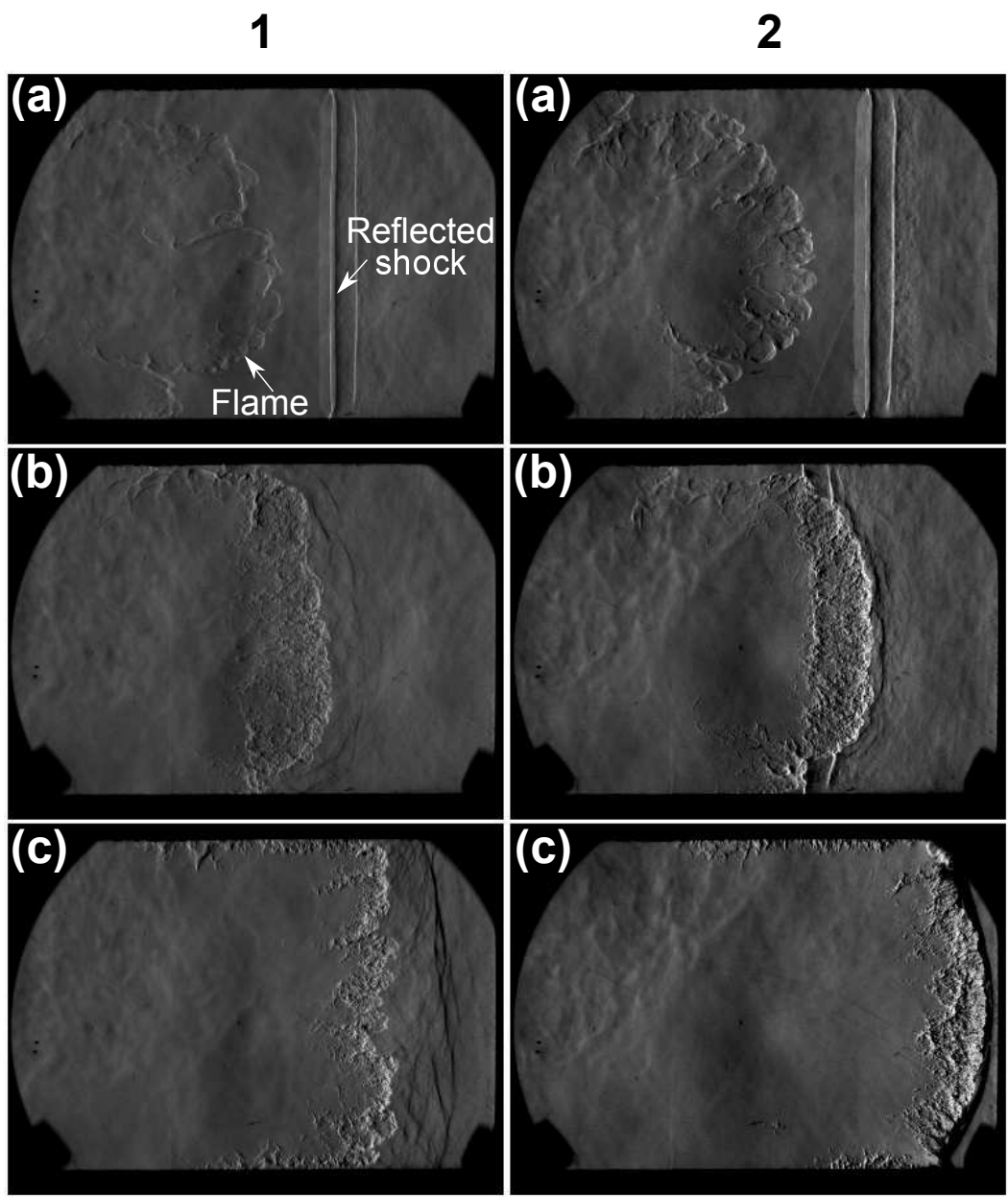}
\end{center}
\caption{The interaction of the reflected shock with a turbulent flame in C$_3$H$_8$+5O$_2$ leads to the flame break-up and its subsequent acceleration; 1) at 4.1 kPa initial pressure, frames shown every 206 $\mu$s apart and (2) and 4.8 kPa, frames shown 129 $\mu$s apart (videos as supplemental material).}
\label{fig:exp26_25wall}
\end{figure}

Figure \ref{fig:exp26_25wall} shows the evolution of the flame resulting from the interaction with the reflected shock at the critical regime, at an initial pressure of 4.1 and 4.8 kPa, for two cases where a transition to detonation was not observed and observed, respectively.  In both cases, the dynamics follow the same sequence.  The first frames show the structure of the wrinkled flame and the bifurcated structure of the reflected shock \cite{mark1958interaction}.  The interaction between the two significantly disturbs the flame structure.  The flame structure is now broken up into smaller individual pockets of non-reacted gas.  The mechanism for this break-up is consistent with the Richtmyer-Meshkov instability, augmented by the 3D structure of the reflected bifurcated shock \cite{khokhlov1999numerical}.  Note also that the compression of turbulent gas ahead of the flame by the reflected shock is also expected to reduce the length and time scales of the turbulence.  

The resulting increase in the surface area between burned and unburned gas augments significantly the burning velocity of the flame, and the flame accelerates forward driving a series of compression waves, which eventually form a shock. In the case of Fig. \ref{fig:exp26_25wall}.2, this acceleration is faster, and the acceleration can proceed in the gas available ahead of the flame.  In this case, the detonation forms near the sides, where the reflection of the non-planar leading shock, inherited by the non-planarity of the flame itself, gives rise to local shock reflections favoring the transition.      

\begin{figure}
\begin{center}
\includegraphics[width=140mm]{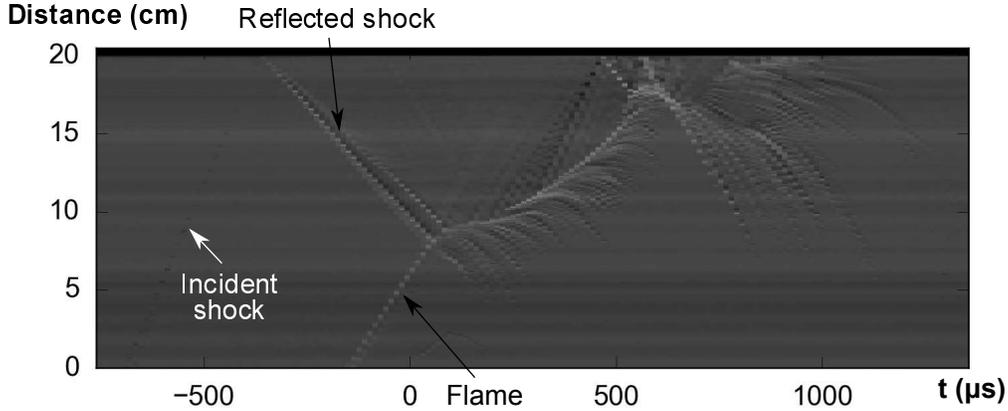}
\end{center}
\caption{A space-time diagram evolution of the interaction of Fig. \ref{fig:exp26_25wall} along the center of the channel illustrating the flame acceleration to the sonic condition.}
\label{fig:exp26wallxt}
\end{figure}

\begin{figure}
\begin{center}
\includegraphics[width=140mm]{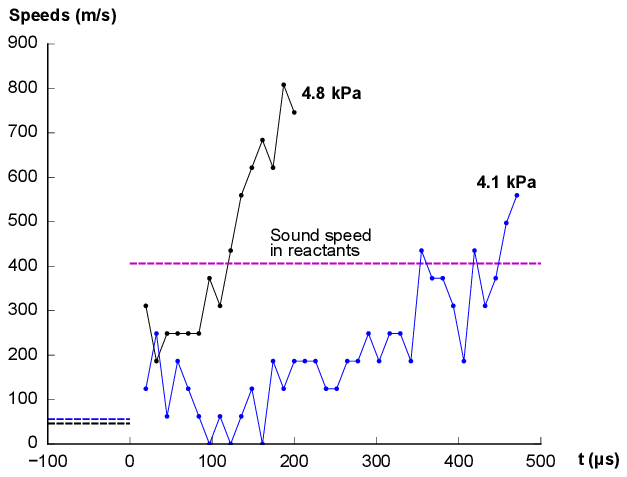}
\end{center}
\caption{The speed of the flames corresponding to Fig.\ \ref{fig:exp26_25wall} after the interaction with the reflected shock; broken lines denote the burning velocity measured prior to the interaction; time 0 corresponds to the first interaction between the flame and the reflected shock.}
\label{fig:exp25_26wallspeeds}
\end{figure}

\begin{figure}
\begin{center}
\includegraphics[width=140mm]{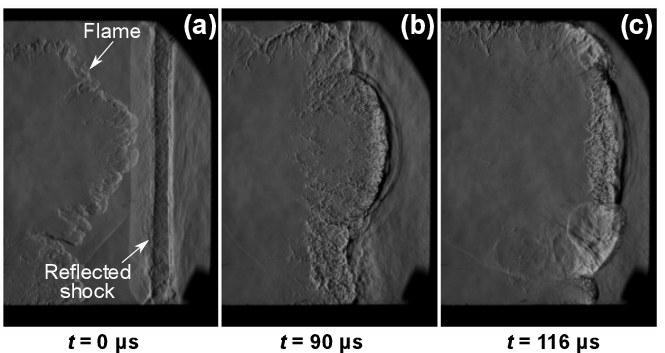}
\end{center}
\caption{The interaction of the reflected shock with a turbulent flame at 6.2 kPa illustrating the onset of detonation from the confluence of shocks driven by the flame acceleration (video as supplemental material).}
\label{fig:exp23wall}
\end{figure}

The acceleration is best seen on the space time diagram of Fig.\ \ref{fig:exp26wallxt}, where the evolution of the experiment at 4.1 kPa is tracked along a central band 3-pixels-wide, and in Fig.\ \ref{fig:exp25_26wallspeeds}.  Evident from the reconstruction of Fig. \ref{fig:exp26wallxt} is the backward acceleration of the gases processed by the flame by the rear facing streaks of the small scale non-reacted pockets and the forward facing acceleration of the flame leading edge.  The acceleration generates forward facing compression waves, seen distinctly in this reconstruction.  The flame path becomes parallel with the forward facing waves, indicating that the flame burning velocity is now comparable with the local sound speed, approximately 410 m/s; this is compatible with the mechanism deduced from the cylinder experiments.  This suggests that the mechanism at play for the acceleration in this case is one for which the burning velocity is increased beyond the sonic condition. This implies that the local pressure increase can no longer be relaxed gasdynamically sufficiently fast, and an internal shock forms ahead of the flame.  This internal shock is subsequently strengthened by further acceleration of the burning rate.  In the experiment of Fig. \ref{fig:exp26_25wall}.2 (see also Fig. \ref{fig:exp25_26wallspeeds}), the shock observed becomes consistent with rapid auto-ignition such that the flame motion can be effected by shock motion itself, i.e., the transition to detonation.  This transition mechanism is similar to the conventional reactivity gradient mechanism.  In the present case, the acceleration of the flame is first due to the enhancement of the burning rate by RM instability, which makes the flame itself in phase with acoustic waves and results in pressure amplification and internal shock formation.   
Experiments at higher pressures yielded significantly higher burning velocities (in excess of 20 times the laminar burning velocity) in the flame following the incident shock, and transition to detonation was observed with time scales less than 100 $\mu$s.  An example is shown in Fig. \ref{fig:exp23wall}.  Here, the transition to detonation was found internally to the disturbed flame brush by the reflection of the internally formed shocks, owing to the initial non-planarity of the leading shock.  This was again consistent with the experiments performed with the cylinder.

\section{Acceleration mechanism and DDT}
\addvspace{10pt}
The picture that emerges from the experiments performed with the interaction of the shock flame complex with a cylinder and a flat wall is that the reflected shock, in both cases, significantly enhances the burning rate of the flame.  The origin of this enhancement is the disruption of the flame structure by the passage of the shock, with auto-ignition being negligibly slow at early times.  The exact mechanism of flame disruption is difficult to assess with certainty.  While the Richtmyer-Meshkov instability may disrupt the original turbulent flame, the reflected shock in all the cases investigated has a bifurcated structure, with a re-circulation bubble.  This structure may also significantly disturb the flame.  Shock-turbulence interaction may also be responsible for the reduction of the smaller scales of the turbulence.  

Once the flame burning velocity approaches the sonic condition in the fresh gases, the flame motion is in phase with pressure waves and a strong shock is eventually driven in front of the flame.  There appears a quasi-steady period at this state.  Subsequent increase in burning rate amplifies the leading shocks, which can trigger auto-ignition, usually through shock reflections. This picture of flame acceleration in very turbulent flames appears to be consistent with the picture borne out from the numerical calculations of Poludnenko et al. \cite{poludnenko2011spontaneous}, who investigated the fate of flames with arbitrarily high turbulent burning velocities.  That such high turbulent burning velocities are possible has been shown in the present study, consistent with previous indirect observations of Khokhlov et al. \cite{khokhlov1999numerical}.

The acceleration time scales determined in the wall experiments at critical conditions correspond approximately to the conditions in the cylinder experiments, suggesting a possible correlation between two time important time scales of the phenomenon:  the flow time associated with gas dynamic cooling as the flame gets convected downstream of the obstacle, and the characteristic time for flame acceleration.  When the characteristic time for flame acceleration is smaller than the flow time of the flame over the obstacle, the flame is expected to have a sufficient time to exploit the amplification of the burning rate by the passage of the shock.  When the time scale for flame amplification is much longer than the flow passage time over the obstacle, the flame does not have sufficient time to accelerate.  This is in accord with the flow fields observed in the experiments where a clear demarcation between regimes of very fast amplification and regimes with negligible amplification was established.   

In conclusion, the mechanism identified as controlling the transition to detonation when a sufficiently strong turbulent flame-shock complex propagates in a congested area is the enhancement of the burning rate by the reflected shocks.  Auto-ignition from the reflected shocks in the current experiments has been found negligible until a much stronger shock is generated by flame amplification.  It thus appears that a criterion for DDT in a congestion giving rise to reflected shocks could likely be established in terms of a time scale for the 1D problem of flame acceleration following the head-on shock-flame interaction.  Future study should focus on experiments and analysis of this canonical problem, to determine if this problem itself admits simple scaling laws.

\section*{Acknowledgements}

M.I.R acknowledges the financial support from both Shell and NSERC via a Collaborative Research and Development Grant, ``Quantitative assessment and modeling of the propensity for fast flames and transition to detonation in methane, ethane, ethylene and propane".

%% References

\bibliographystyle{elsarticle-num-CNF}
\bibliography{references}






% \end{thebibliography}

\end{document}